\documentclass[Afour,sagev,times]{sagej}

\usepackage{moreverb,url}
\usepackage{array,ragged2e}
\usepackage{threeparttable}
\usepackage[labelfont=bf,labelsep=period,justification=raggedright]{caption}
\usepackage{multirow}
\usepackage{adjustbox}
\usepackage{float}
\usepackage[english]{babel}
\usepackage{bm}
\usepackage[colorlinks,bookmarksopen,bookmarksnumbered,citecolor=red,urlcolor=red]{hyperref}
\newcolumntype{L}[1]{>{\RaggedRight\hspace{0pt}}p{#1}}
\def\volumeyear{2022}

\begin{document}

\runninghead{Cherlin and Wason}

\title{Utilising high-dimensional data in randomised clinical trials: a review of methods and practice}

\author{Svetlana Cherlin\affilnum{1}, Theophile Bigirumurame\affilnum{1}, Michael J Grayling\affilnum{1}, \\J\'er\'emie Nsengimana\affilnum{1}, Luke Ouma\affilnum{1}, Aida Santaolalla\affilnum{2}, \\ Fang Wan\affilnum{3}, S Faye Williamson\affilnum{1}, James M S Wason\affilnum{1}}

\affiliation{\affilnum{1}Population Health Sciences Institute, Newcastle University, Newcastle upon Tyne, UK\\
\affilnum{2}Translational Oncology \& Urology Research Group, Centre for Cancer, Society \& Public Health, King's College London, UK\\
\affilnum{3}Department of Mathematics and Statistics,  Lancaster University, Lancaster, UK
}

\corrauth{Svetlana Cherlin, Population Health Sciences Institute, Newcastle University, Ridley Building 1, Queen Victoria Road, Newcastle upon Tyne, NE1~7RU, UK.}

\email{svetlana.cherlin@newcstle.ac.uk}

\begin{abstract}
{\bf Introduction:} 
Even in effectively conducted randomised trials, the probability of a successful study remains relatively low. With recent advances in the next-generation sequencing technologies, there is a rapidly growing number of high-dimensional data, including genetic, molecular and phenotypic information, that have improved our understanding of driver genes, drug targets, and drug mechanisms of action. The leveraging of high-dimensional data holds promise for increased success of clinical trials.
 
{\bf Methods:} We provide an overview of methods for utilising high-dimensional data in clinical trials.
We also investigate the use of these methods in practice through a review of recently published randomised clinical trials that utilise high-dimensional genetic data. The review includes articles that were published between 2019 and 2021, identified through the \textit{PubMed} database. 

{\bf Results:} Out of 174 screened articles, 100 (57.5\%) were randomised clinical trials that collected high-dimensional data. The most common clinical area was oncology (30\%), followed by chronic diseases (28\%), nutrition and ageing (18\%) and cardiovascular diseases (7\%).  The most common types of data analysed were gene expression data (70\%), followed by DNA data (21\%). The most common method of analysis (36.3\%) was univariable analysis. Articles that described multivariable analyses used standard statistical methods.  Most of the clinical trials had two arms. 

{\bf Discussion:} 
New methodological approaches are required for more efficient analysis of the increasing amount of high-dimensional data collected in randomised clinical trials. We highlight the limitations and barriers to the current use of high-dimensional data in trials, and suggest potential avenues for improvement and future work.
\end{abstract}


\maketitle

\section{Introduction} \label{sec:intro}
Randomised controlled trials (RCTs) are the gold standard for assessing the safety and efficacy of an experimental treatment. However, despite the growing cost and time associated with  developing and evaluating drugs, the probability of success of RCTs is relatively low.\citep{wong:siah:2019} One of the reasons is that there is rarely a ``one size fits all" approach in most clinical areas because treatment typically has a heterogeneous effect on patients with different pathogenic mechanisms. For example, in a study that investigated predictors of response to Methotrexate in early rheumatoid arthritis, \citep{saevarsdottir:etal:201} 75\% of the patients experienced a good response rate, according to the EULAR response criteria.\citep{vanGestel:etal:1995} This study found that several demographic and clinical characteristics (including age, sex, smoking status and symptom duration) are associated with response to Methotrexate.  Subsequently, a double-blind phase IV clinical trial in patients with rheumatoid arthritis identified genetic markers that could partly explain the heterogeneity of response to Methotrexate. \citep{aslibekyan:etal:2014}

With recent advances in the next-generation sequencing technologies, there is a rapidly growing number of human molecular biomarkers that could inform drug mechanisms and increase the success of clinical trials. \citep{nelson:etal:2015} Molecular biomarkers are measurable molecular characteristics (small molecules) that could identify relatively homogeneous disease subsets in terms of clinical features, diagnosis, prognosis, or response to treatment. With the advent of personalised medicine, molecular biomarkers are gaining importance in clinical research.\citep{buyse:etal:2011} The most common types of molecular biomarkers are genomic biomarkers such as deoxyribonucleic acid (DNA) and ribonucleic acid (RNA). Single nucleotide polymorphisms (SNPs), which are the most abundant type of genetic variation, represent a difference in a single nucleotide.  SNPs are often measured (genotyped) across the genome, and the associations between genome-wide SNPs and different human traits, i.e.\ genome-wide association studies (GWAS), are extensively used in genetics.\citep{visscher:etal:2017} GWAS to date have analysed hundreds of thousands of genetic variants generated by next-generation sequencing technologies. 

Proteomics and metabolomics also play an important role in many medical applications and are being increasingly used in drug research and development.\citep{wishart:2016} Proteomics is a study of molecules in proteins that allows characterisation of protein structure and function. Protein biomarkers are also increasingly used in clinical trials in patient stratification, disease diagnosis, and prognosis.\citep{he:2019} Another commonly used genetic biomarker is gene expression, which is a process that regulates the amount of protein or other molecules expressed by the cell, and thus is measured by the amount of the molecules or protein. The advantage of microarray technology is to allow for gene expression profiling which consists of measuring levels of thousands of genes. Changes in gene expression can reflect the change in a cell's environment, such as disease state,\citep{emilson:etal:2008} response to treatment \citep{geeleher:etal:2014} or treatment side effect.\citep{duffy:etal:2020} 
Metabolites, small molecules produced by the body when it breaks down food or drugs, are useful for biomarker discovery because they can be utilised to examine the underlying biochemical activity of cells. Modern technologies, such as mass spectrometry, allow for a large number of metabolies to be measured thus creating a metabolomic profile. \citep{shaham-niv:etal:2021} Metabolic changes are informative of the response to treatment and therefore have the potential to be useful in clinical trials.\citep{kaddurah-dauok:etal:2015} For example, a randomised placebo-controlled clinical trial that examined the effect of sertraline on major depressive disorder patients found that baseline metabolic signatures could be predictive of response or non-response to sertraline.\citep{kaddurah-dauok:etal:2011} 

In clinical trials, biomarkers serve multiple purposes, such as prognosis of the likely progression of a disease, and prediction of the likely clinical outcome.\citep{antoniou:etal:2016} Prognostic biomarkers are those that are associated with disease prognosis in the absence of treatment or in the presence of a standard of care treatment. Predictive biomarkers are those that are associated with the effectiveness of a specific treatment. Predictive biomarkers could be used to identify subsets of patients who are likely to respond to treatment. For example, a pooled analysis of randomised trials  found that women whose breast tumours have overexpressed the human epidermal growth factor receptor 2 ($HER2$) protein or amplified $HER2$ gene ($HER2$-positive) benefited from adjuvant treatment with anthracyclines, while women with $HER2$-negative breast tumours derived no added benefits from adjuvant chemotherapy with anthracyclines.\citep{gennari:etal:2008} Thus, the $HER2$ status of a breast tumour is a predictive biomarker for response to adjuvant treatment with anthracyclines. Prognostic and predictive biomarkers are usually measured once, before the start of treatment. Biomarkers that are measured repeatedly during the trial could be used as a surrogate endpoint, i.e.\ as a proxy for a clinical endpoint. Biomarkers based on a continuous single gene measurement can be used as  classifiers by considering a threshold,  or a series of thresholds,   to specify a biomarker-positive and biomarker-negative group \citep{jiang:etal:2007, simon:2008}. However, identifying single-gene biomarkers requires knowledge and biological interpretation of the disease pathway, which may not always be available.

Recent advances in whole genome biotechnology allow for measuring multiple genetic variants during clinical trials.\citep{hu:dignam:2010, johnson:etal:2014, wanga:etal:2015, wei:etal:2022} This allows biomarkers across multiple genes to be developed, i.e.\ biomarkers based on high-dimensional data. A variety of predictive and prognostic biomarkers based on high-dimensional molecular profiling have been proposed in oncology.\citep{theilhaber:etal:2020, ye:etal:2019, li:etal:2021} These biomarkers are especially relevant for finding potential responders to a treatment in settings where an assay for identifying biomarker-positive patients is not yet available.\citep{freidlin:simon:2005} While prognostic biomarkers based on high-dimensional data are becoming increasingly available, predictive biomarkers based on high-dimensional data are rare due to the challenge of understanding a treatment's mechanism of action.\citep{simon:2014} Additional challenges of using high-dimensional data are identifying which biomarkers to include in the model, and how to effectively/appropriately combine the individual biomarkers.\citep{johnstone:totterington:2009}

In this paper, we provide an overview of several statistical methods for utilising high-dimensional data in the analysis of RCTs. We also present a review of recently published clinical trials that utilised high-dimensional data to investigate how often various methods have been used in practice.  

\section {Overview of methods for utilising high-dimensional data in clinical trials} \label{sec:overview}
In this section, we describe statistical methods used for analysing  high-dimensional data in RCTs; many of which have been implemented in standard statistical software such as R. \citep{rcore:team:2021} A summary of these methods is provided in Table \ref{tab:methodsummary}. When considering suitability of the methods, it is important to distinguish between testing for association and prediction. Association tests, such as the Chi-Square test, can shed light on the biological processes by providing better understanding of the phenomenon in question. Association tests are useful for testing hypotheses about the differences between the groups of observations, such as the difference between the treatment arms, or for finding biomarkers that are associated with response to treatment.

In prediction analysis, statistical models such as regression are applied to data in order to build predictors that could be applied to future studies. The quality of prediction should be assessed on an independent dataset using some measure of the discrepancy between  the observed and  predicted outcomes. 

Some of the methods we review in this manuscript focus on either testing for association or on prediction, while others focus on both. However, it is important to note that models that have high power to detect associations do not necessarily have high predictive power.\citep{shmueli:2010} 

\subsection{Notation} \label{sec:notation}
In this section we describe a two-arm RCT where participant $i$ ($i=1,\dots, n$) is randomised to either an intervention arm ($t_i = 1$) or control arm ($t_i=0$). For each participant $i$, a set of $j=1,\dots,m$ biomarkers, $x_{ij}$, are collected, and an outcome $y_i$ is measured. Regression modelling is often used to model the outcome $y_i$ as a function of the covariates $x_{ij}$, which are measurable quantities related to the outcome. For different types of  outcome, different types of regression are used. The most common types are linear regression (for continuous outcomes), logistic regression (for binary outcomes) and Cox regression (for time-to-event outcomes).  Linear regression models the mean of the continuous outcome, assuming that the outcome is normally distributed. Logistic regression models the log odds, $\mathrm{logit}(p_i) =  \mathrm{log} \left(  \frac{P(Y_i=1)}{1-P(Y_i=1)} \right)$, where $P(Y_i=1)$ denotes the probability of a successful outcome. Cox regression models the hazard ratio of an event at time $t$, $\mathrm{log}\left(  \frac{h_i(t)}{h_{0i}(t)}  \right)$, where $h_{0i}(t)$ is the baseline hazard at time $t$. In these regression models, the link function of the response variable connects the covariates with the expected value of the outcome variable in a linear way, while the covariates are being weighted by their coefficients. The null hypothesis of a specific coefficient being zero represents testing for an effect of the corresponding covariate.

\begin{table}[h!] 
	\centering
	\begin{threeparttable}
		\twocolumn[%
		\maketitle
		\noindent
		\caption{Summary, advantages and disadvantages of methods utilising high-dimensional data. 
			\label{tab:methodsummary}}
		\begin{tabular}{ L{100pt} | L{180pt} | L{100pt} | L{100pt} }
			\toprule
			Method & Summary & Advantages & Disadvantages \\
			\midrule
			Univariable approach &Testing one biomarker at a time & Simplicity. &  Multiple testing issue\\  
			Multivariable approach &Testing a number of biomarkers simultaneously & Fitting a single model for a several biomarkers & Overfitting   \\ 
			Penalised approach & 
			Penalises regression coefficients, causing them to shrink, maybe to zero  & Prevention of overfitting & Tuning of parameters \\  
			Random forests & Collection of regression or classification trees & Allows modelling non-linear interactions & Lack of intuitive interpretation\\  
			Support vector machines &Building a classifier by fitting a hyperplane between different groups of observations &Allows modelling non-linear interactions& Computational complexity\\ 
			Cluster analysis &Grouping data based on a measure of similarity & Allows modelling non-linear interactions & Sensitivity to outliers\\ 
			Gene sets and networks & Undirected graphs representing associations between the genes &  
			Dimensionality reduction & Computational complexity \\ 
			Principal component analysis & Transforming high-dimensional data into low-dimensional variables that account for most of the original data's variation  &  Allows modelling non-linear interactions & Lack of intuitive interpretation of the principal components\\ 
			Adaptive signature design & Constructing a low-dimensional score from high-dimensional data & Finding group of patients benefiting from  treatment & Multiple testing issue 
		\end{tabular}
		]
	\end{threeparttable}
\end{table}

\subsection{Univariable approach} \label{sec:univar}
A univariable approach consists of testing a single biomarker's relationship to a response variable.  
In linear regression, the outcome $y_i$ for patient $i$ takes the form
\begin{equation*}
y_i = \beta_{j0} + \beta_{j1}t_{i} + \beta_{j2}x_{ij} + \beta_{j3}t_ix_{ij} +\epsilon_i,
\end{equation*}
where $\epsilon_i \sim N(0, \sigma^2)$ is the error term. In logistic regression, the probability of the outcome $y_i$ for patient $i$ takes the form:
\begin{equation*}
\mathrm{logit}(p_i) = \beta_{j0} + \beta_{j1}t_{i} + \beta_{j2}x_{ij} + \beta_{j3}t_ix_{ij}.
\end{equation*}
In Cox regression, the hazard ratio of an event at time $t$ for patient $i$ takes the form:
\begin{equation*}
\mathrm{log}\left(  \frac{h_i(t)}{h_{0i}(t)}  \right) = \beta_{j1}t_{i} + \beta_{j2}x_{ij} + \beta_{j3}t_ix_{ij}.
\end{equation*}

The null hypothesis $H_{j2}: \beta_{j2}$ = 0 represents testing for a prognostic effect of biomarker $j$, while the null hypothesis $H_{j3}: \beta_{j3}$ = 0 represents testing for a predictive effect of biomarker $j$. These hypotheses could then be tested using a Wald test, for example.

Applying statistical tests to one biomarker at a time could result in an inflated number of false positives, due to multiple independent comparisons.\citep{herzog:etal:2019} To prevent this, the Bonferroni correction \citep{bland:altman:1995} is often applied, which adjusts the significance level of individual tests to level $\alpha/m$, where $m$ is the number of tests and $\alpha$ is the desired family-wise error rate. To reduce multiple testing burden, a two-step procedure has been proposed \citep{wang:etal:2021} that  accounts for correlation between the biomarkers via penalised regression. In the first stage of the procedure, a screening test selects a subset of biomarkers, and in the second stage, only the selected biomarkers are tested for interaction. 

An additional challenge in detecting interactions is due to the large sample size required to obtain high power.\citep{brankovic:etal:2019, brookes:etal:2004} In the case of a binary biomarker, in which the trial population can be divided into biomarker-positive and biomarker-negative subgroups, the sample size for testing a null hypothesis of no interaction is at least four times higher than the sample size needed to test the main effect (see Appendix). 

Univariable analysis models are straightforward to fit and produce intuitive results. However, in the real word there is often more than just one biomarker involved. Analysing one biomarker at a time ignores the correlation between the biomarkers, which could lead to incorrectly concluding that some biomarkers are predictive.

\subsection{Multivariable approach} \label{sec:multivar}
A multivariable  regression takes into account two or more biomarkers. Similarly to the univariable regression, there are three commonly used regression types: linear (for continuous outcomes), logistic (for binary
outcomes) and Cox regression (for time-to-event outcomes), which take the following form when $m$ biomarkers are simultaneously adjusted for:
\begin{equation*}
y_i = \beta_{j0} + \beta_1t_i +  \sum_{j=1}^{m} \beta_{j2}x_{ij} + \sum_{j=1}^{m}\beta_{j3}t_ix_{ij} + \epsilon_i,
\end{equation*} 
\begin{equation*}
\mathrm{logit}(p_i) = \beta_{j0} + \beta_1t_i +  \sum_{j=1}^{m} \beta_{j2} x_{ij} + \sum_{j=1}^{m}\beta_{j3} t_i x_{ij}\mathrm{,~and}
\end{equation*}
\begin{equation*}
\mathrm{log}\left(  \frac{h_i(t)}{h_{0i}(t)}  \right) =  \beta_1t_i +  \sum_{j=1}^{m} \beta_{j2} x_{ij} + \sum_{j=1}^{m}\beta_{j3} t_i x_{ij},
\end{equation*}
respectively. Multivariable analysis estimates the contribution of each biomarker $x_{ij}$ while adjusting for the effect of other biomarkers or covariates. Therefore, unlike univariable analyses, it takes into account correlation between biomarkers.

The main drawback of the multivariable approach is the large number of parameters that may be included. With high-dimensional data, this approach can lead to a model with more  parameters  than observations (i.e.\ the ``curse of dimensionality"). In this case, multivariable linear regression cannot be used because the unique ordinary least squares estimators of the regression coefficients are not defined. To reduce the complexity of the model, several variable selection approaches have been proposed, including machine learning approaches (discussed below). However, a large number of parameters in the model could still lead to overfitting, which is the phenomenon of modelling the observed data too precisely so that it captures the noise in the data. In this case, the model shows an inferior performance when applied to a new dataset. To reduce the potential effects of overfitting, a rule-of-thumb is that at least ten events are required per variable in logistic and Cox regression models, though this rule is often debated.\citep{grant:etal:2019} For linear regression estimated using ordinary least squares, the number of covariates that can be included in the model is generally higher; it has been shown that two subjects per value would be sufficient for adequate estimation of regression coefficients.\citep{austin:steyerberg:2015}

\subsection {Regularised (penalised) regression} \label{sec:penalised}
Regularised, or penalised, approaches penalise models by  shrinking the estimates of the regression coefficients. Suppose a regression model with a  $(m+1)$-dimensional vector of covariates $\boldsymbol{\beta} = (\beta_0, \beta_1, \dots, \beta_m)^T$  is fitted by maximising the log-likelihood function $\ell(\boldsymbol{\beta})$. In penalised regression, $\ell(\boldsymbol{\beta})$ is maximised subject to a penalty function $P(\boldsymbol{\beta})$ and a regularisation parameter $\lambda$, that is, $\boldsymbol{\hat\beta} = \mathrm{argmax}[\ell (\boldsymbol{\beta}) - \lambda P(\boldsymbol{\beta})].$ As a result, the regression coefficient estimate $\boldsymbol{\hat\beta}$ is shrunk towards zero in comparison to the maximum likelihood estimate, with $\lambda$ controlling the amount of shrinkage.

The method induces different degrees of sparsity, depending of the type of penalty used. For example, the Least Absolute Shrinkage and Selection Operator (LASSO) regression \citep{tibshirani:1996} allows shrinkage of the coefficients to zero by penalising the model with $P(\boldsymbol{\beta}) = ||\boldsymbol{\beta}||_{\ell_1} = \sum_{j=1}^{m}|\beta_j|$ and is therefore a sparse method which allows for variable selection. Another type of penalised regression is ridge regression \citep{cessie:how:1992} in which the penalty function has the form $P(\boldsymbol{\beta}) = ||\boldsymbol{\beta}||_{\ell_2} = \sum_{j=1}^{m}\beta_j^2$. Ridge regression shrinks the coefficients \textit{towards} zero, however it does not shrink them to zero. Elastic net  \citep{zou:hastie:2005} is a type of  penalised regression in which both penalties are used, i.e.\
\begin{equation*}
\boldsymbol{\hat\beta} = \mathrm{argmax} \left[  \ell (\boldsymbol{\beta}) - \lambda \left( \eta \sum_{j=1}^{m}|\beta_j| + \frac{1-\eta}{2} \sum_{j=1}^{m}\beta_j^2 \right) \right].
\end{equation*}
The combination of the penalties is controlled by a penalty weight parameter $\eta$. When $\eta = 1$, the elastic net is identical to LASSO, whereas when $\eta = 0$ it is identical to ridge. 
Elastic net combines setting of the coefficients to zero using LASSO and shrinking of the coefficients using ridge, to improve the model's performance.
A penalised logistic regression model, which included ten genes, was used  to predict the  overall complete pathologic response rate in a phase II genomic study of ixabepilone as neoadjuvant treatment for breast cancer.\cite{baselga:etal:2009} A pharmacogenetic study used ridge regression to predict a response to treatment.\citep{geeleher:etal:2014} It has been found that using LASSO regression improved the accuracy of the treatment effect estimator in a RCT.\cite{bloniarz:etal:2016} A review of neoadjuvant clinical trials in breast cancer that analysed gene expression data \citep{ternes:etal:2014} found that penalised methods outperform competing methods when applied to estrogen receptor-positive (ER+) early breast cancer patients treated with neoadjuvant aromatase inhibitor letrozol. However, an application of a  penalised high-dimensional Cox model to an early breast cancer RCT of chemotherapy with or without adjuvant trastuzumab resulted in highly variable expected survival probabilities  with very large confidence intervals.\citep{ternes:etal:2017}

Group-lasso \citep{yuan:lin:2006} is a special case of LASSO that performs selection of important groups of variables.  For example, the groups could represent specific biological pathways of the biomarkers, or variables that reflect a specific aspect of a treatment. Extending the group-lasso by considering interactions,\cite{lim:hastie:2015} however, can result in many false positive interactions for high-dimensional problems.

Penalised regression requires optimisation of the penalty parameter, which could be done using cross-validation. In the cross-validation procedure, a model is fitted to a subset of the data and its accuracy is assessed on a different subset of the data. The process is repeated multiple times with different partitions of the data for fitting (training subset) and assessing (testing subset). Parameters that lead to the best accuracy are chosen. However, when  cross-validation is used to examine model performance, tuning of the parameters requires nested cross-validation, in which the inner cross-validation (for tuning of parameters) is encapsulated inside the outer cross-validation (for assessing model performance). This procedure requires large sample sizes. It is also necessary to ensure homogeneous partitioning of the data with respect to important features, in order to achieve a valid cross-validation procedure \cite{krstajic:etal:2014}.

\subsection{Machine learning approaches} \label{machlearn}
Machine learning is a class of algorithms that analyse data based on existing (training) data.\citep{jordan:mitchell:2015} Machine learning algorithms can either be supervised or unsupervised, with the difference being the labelling of the input data. In supervised machine learning algorithms such as classification, the training data is labelled, while in unsupervised methods such as clustering, the training data is not labelled. Supervised approaches are used for predictive modelling when the classification of the training data is known in advance, and the trained algorithm is used to predict or classify new data with unknown classification, such as response or non-response to treatment in clinical trials. Unsupervised methods are used for feature selection problems, such as identifying a predictive biomarker in the context of biomarker analysis,  and dimensionality reduction \cite{weissler:etal:2021}. 

\subsubsection{Random forests} \label{randomforests}
Random forests  are a type of high-dimensional nonparametric model aimed at prediction,\citep{brieman:2001} and therefore belong to the class of supervised machine learning algorithms. They are represented as a collection of regression trees (for a continuous outcome) or classification trees (for a binary outcome). Each tree is a decision model that consists of a recursive partitioning of a dataset into subsets that are determined by a randomly selected group of input variables. The subsets are homogeneous with respect to the group of variables. At each node of a tree, different groups of variables might be used. Random forests are formed by trees constructed from training datasets sampled with replacement from the original dataset. The remaining samples form the testing datasets and are used for assessing prediction accuracy. For example,  the probability of misclassifying an observation could be used as a measure of prediction accuracy. Random forests are flexible in that regression and classification trees can incorporate non-linear interactions between the variables.\citep{reif:etal:2009}

Traditional random forests are designed for one treatment group and are therefore suitable for prognostic, rather than predictive, purposes. A few adaptations of the method for more than one treatment group have been developed that facilitate identification of a subset of patients who benefit from the treatment. For example, the ``Virtual Twins" method \citep{foster:etal:2011} is a random forest-based method of identifying a subgroup of enhanced treatment effect by incorporating treatment-covariate interactions.

A variation of the random forest has been developed, that uses a measure based on a difference in survival times as an alternative to the accuracy prediction, for deciding on a best possible split.\cite{ubels:etal:2020} When applied to a phase III RCT with high-dimensional SNP data, this approach has been shown to outperform a univariable analysis. The challenges of this method include specifying model parameters, such as the number of trees in the forest. 

\subsubsection{Support vector machines }
Support vector machines (SVM) are a supervised machine learning method for building a classifier that can be used to account for non-linear relationships between variables.\citep{vapnik:1995} SVM assign an observation to a specific category, or class, by fitting a hyperplane between the samples from different classes so that the distance between the hyperplane to the nearest sample is maximised. This distance is maximised using support vectors, i.e.\ data points that are closer to the hyperplane. SVM involve transforming the data using a kernel function to allow linear separation of the data. An advantage of SVM is that it can effectively incorporate high-dimensional data that can be noisy and/or correlated. It has been widely applied to classification problems using high-dimensional biomarkers.\citep{ hua:sun:2001, dror:etal:2005, liu:etal:2006, ng:mishra:2007, huang:etal:2018} SVM could be used in RCTs if treatment-covariate interaction effects are introduced into the feature space of SVM. Using SVM constructed from the combination of brain imaging and demographic and clinical biomarkers, a group of Mild Cognitive Impairment patients who were most likely to cognitively decline has been identified.\cite{kohannim:etal:2011} Limitations of SVM include their computational complexity, especially the need to optimise their parameters. 

\subsection {Cluster analysis} \label{sec:clust}
Clustering methods are unsupervised methods of grouping  data based on some measure of similarity, so that the observations in each group are similar (but dissimilar to those in other groups). The most common measure of similarity between the observations is correlation. Traditional clustering methods include hierarchical clustering and partitioning.\citep{reynolds:etal:2006} In hierarchical clustering, the data is organised into a tree-shape structure (a dendogram) constructed from hierarchical series of nested clusters, while partitioning does not assume hierarchical relationships between clusters. An example of partitioning is $k$-means clustering, which partitions the data into a pre-specified number $k$ of mutually exclusive groups so that the the sum of the squared distances between the members of the group and the means of the clusters is minimised.\citep{hartigan:1975} Another example is Partitioning Around Medoids clustering, which is similar to the $k$-means but is more robust to outliers.\citep{kaufman:rousseeuw:2009} 

Hierarchical clustering employs agglomerative and divisive strategies.  Hierarchical agglomerative clustering starts by treating each sample as a separate cluster and then merges the most similar clusters together. This process is repeated iteratively until all samples are clustered. Hierarchical divisive clustering  starts by treating all the observation as one cluster, and them recursively splits the cluster into two, until the desired number of clusters is obtained. Hierarchical clustering could be used to analyse genes that are differentially expressed between different experimental conditions, such as the different treatment groups in clinical trials. To estimate the number of clusters in the dataset, consensus clustering could be used which utilises bootstrapping to classify each observation multiple times. Finally, observations are assigned to the cluster with the highest consensus score and the number of clusters is derived from objective metrics.\citep{monti:etal:2003} Other methods, called model-based clustering, exploit the same idea of making clustering robust to model misspecification and estimation of the number of clusters. They assume that observations follow a mixture of distributions rather than belonging to discrete classes. 

Clustering is often used in gene expression analysis because it simplifies visualisation and allows one to trace specific biological pathways.\citep{jiang:etal:2004, dhaeseleer:2005, vavoulis:etal:2015, oyelade:etal:2016} Moreover, it can be used to identify specific disease subtypes. For example, hierarchical clustering was able to identify pre- and post-vaccine samples in a study of the effect of an influenza vaccination on gene expression. \citep{drury:etal:2019}

\subsection {Gene sets and networks}
Gene networks belong to the class of the unsupervised machine learning algorithms. They are undirected graphs with nodes representing genes and edges representing gene-gene associations. Genes with similar co-expression patterns are then grouped into modules using clustering techniques. Different types of co-expression networks are discussed elsewhere.\cite{zhang:horvath:2005, vanDam:etal:2018} 

Weighted gene co-expression network analysis (WGCNA) \citep{langfelder:horvath:2008} is a common co-expression network method that is used for finding clusters of highly correlated genes. It summarises the clusters using the representative gene (the eigengene), thus performing dimensionality reduction. The eigengene is a vector that represents the expression of all the genes in the model. WGCNA has been used to analyse metabolites in an ancillary study of vitamin D supplementation for the prevention of asthma.\cite{lee-sarwar:etal:2019} The eigenvalues for the modules of metabolites were used to find association with asthma.

Gene set enrichment (GSE) is another subtype of gene networks that clusters genes into pre-defined sets that share common biological functions, and summarises the gene expressions into a single score for each set. Scores represent the extent of the differences in gene expression between the phenotypic classes of interest, for example tumours that are responsive or non-responsive to treatment. Testing the statistical significance of the scores allows detection of an enrichment signal.\citep{subramanian:etal:2005} GSE analysis has been used to compare advanced colorectal cancer subtypes in a RCT of first-line treatment of metastatic colorectal cancer.\citep{takahashi:etal:2021} Gene networks are useful for dimensionality reduction of a large number of correlated genes. To our knowledge, this method has not been used for comparing treatment arms or finding predictive biomarkers in clinical trials.

\subsection{ Principal component analysis} \label{sec:pca}
Principal component analysis (PCA) is a statistical technique that provides information on directions of variability in  data. PCA consists of transforming  high-dimensional data into a lower-dimensional set of variables (principal components) such that the first principal component (PC) is associated with the largest source of variation, the second PC with the largest remaining source of variation and so on. The procedure of computing the PCs involves computing the eigenvalues and eigenvectors for the covariance matrix of standardised data. PCs are formed by transforming the original data using a matrix constructed from the eigenvectors.\citep{jolliffe:cadima:2016}

Each PC is constructed as a linear combination of the original high-dimensional data in such a way that the PCs are mutually uncorrelated. Thus, PCs could prevent multicollinearity issues in regression models and be very useful for correlated biomarkers. Once computed, the PCs can be used as covariates in linear regression models, as well as a dimensionality reduction technique for clustering. PCA also makes high-dimensional data more suitable for visualisation.  For example, PCs are widely used to identify genetic variation associated with geographic region, \citep{abegaz:etal:2019} with most geographic variation explained by the first two PCs. However, it has been shown that in the analysis of gene expression data, many more PCs might be needed to detect relevant variability, depending on the sample sizes and effect sizes. \citep{lenz:etal:2016}

A challenge of PCA is the interpretation of the PCs, as well as identifying the most informative PCs. In the field of clinical trials, the use of PCA is limited to finding prognostic rather than predictive biomarkers. 

\subsection {Adaptive signature design and risk scores} \label{asd}
Adaptive signature design methods utilise high-dimensional data to construct a low-dimensional (or scalar) signature. They combine information from multiple genetic markers to create a signature that could be used for diagnostic, prognostic or predictive purposes. Adaptive signatures are motivated by the fact that genetics play an important role in the heterogeneity of disease progression and response to treatment, and could therefore be used to facilitate personalised medicine. The original adaptive signature design constructed a low-dimensional signature based on the interaction between the treatment and the high-dimensional baseline biomarker data. \citep{freidlin:simon:2005, freidlin:etal:2010} It was developed for situations with no pre-defined predictive biomarker and utilised a threshold on the number of biomarkers included in the signature. Initially, two non-overlapping groups of trial participants have been used to develop and validate the signature,\cite{freidlin:simon:2005}  while later a cross-validation has been implemented, which uses patient information more efficiently. \cite{freidlin:etal:2010} 

A few studies \cite{radmacher:etal:2002,matsui:etal:2012,cherlin:wason:2020,cherlin:wason:2021b} construct a signature as a sum of the effects of the interactions between the treatment and each of the covariates separately. In these methods, the adaptive signature is represented by a single score for each patient. Specifically, for a binary outcome, a single covariate logistic model is fitted for each biomarker $j = 1,\dots,m$ as follows:
\begin{equation*}
\mathrm{logit}(p_i) = \beta_0 + \beta_1 t_i + \beta_{j2} x_{ij} + \beta_{j3} t_i x_{ij},
\end{equation*}
where $p_i$ is the probability of the outcome of interest, $\beta_{j2}$ represents a prognostic effect of  biomarker $j$,  and $\beta_{j3}$ represents a predictive effect of biomarker $j$.

A risk score for patient $i$ ($RS_{i}$) is computed as the sum of the maximum likelihood estimate of the treatment-covariate interaction coefficients $\hat \beta_{j3}$ weighted by the value of the biomarker $x_{ij}$, i.e.\
\begin{equation*}
RS_i = \sum_{j=1}^{m} \hat \beta_{j3} x_{ij}.
\end{equation*}
The collection of risk scores $RS_i$ for all $i$ could be subdivided in different ways to represent different strata of patients in terms of the predicted treatment benefit. \citep{cherlin:wason:2020, cherlin:wason:2021b}  At the end of the trial, a test is performed for the overall comparison between the arms, as well as for the comparison between the arms in the subgroup, using an $\alpha$-splitting approach to control the type I error rate.
Alternatively, they could be used as covariates to test for an association with the outcome.\citep{matsui:etal:2012}

Adaptive signature designs often utilise a combination of the previously described approaches. For example, an adaptive signature which is predictive of response to MAGE-A3 immunotherapeutic in patients with metastatic melanoma has been developed and validated in a randomised phase II trial \citep{ulloa-montoya:etal:2013} using a variation of PCA  and hierarchical clustering. Another phase II trial \cite{dreno:etal:2018} used a combination of a scoring system and the penalised approach.  For each patient $i$, the following  risk score was constructed that represented the hazard ratio under the two treatments on the logarithmic scale: 
\begin{equation*}
RS_i = \mathrm{log}[h_0 (t|\bm{x}_i)]-\mathrm{log}[h_1 (t|\bm{x}_i)],
\end{equation*}
where $h_j(t|\bm{x}_i)$ is the hazard rate for treatment $j = \{0,1\}$ for patient $i$, and  $\bm {x}_i$ is the vector of gene expressions for patient $i$. The hazard functions were estimated with penalised Cox regression. 

The adaptive signature designs are applied in a post-hoc manner, i.e.\ they identify the subgroup of patients at the end of the trial and therefore do not fit the classical definition of an adaptive design. Rather, they are adaptive in the sense that they allow adaptive selection of patient subgroups. For example, an adaptive signature design has been proposed that finds the optimal subgroup in terms of maximising the power for identifying treatment benefit. \cite{zhang:etal:2017} To address the issue of adaptive changes in trials, the risk scores-based adaptive signature has been utilised in the adaptive enrichment framework, where the trial population is adaptively enriched with patients who are predicted to benefit from the treatment.\cite{cherlin:wason:2021a}

In summary, adaptive signature designs have the advantage of improving the efficiency of clinical trials by identifying enhanced benefit subgroups. More reliably identifying patient subgroups who benefit from the treatment would prevent the situation in which a potentially effective treatment is disregarded because the treatment effect  in the overall population is overlooked. Moreover, adaptive signature designs have the potential to avoid patients who receive no benefit from receiving the treatment, thus preventing unnecessary exposure to possible side effects. However, adaptive signature designs come with a statistical challenge of a multiple comparisons issue. Additionally, there may be a need for dimensionality reduction in situations with a large number of baseline biomarkers. \citep{bhattacharyyaa:rai:2019}

\section{Current use of  methods for utilising high-dimensional data in RCTs} \label{sec:current_use}
\subsection{Review methods}
We performed the following literature search of RCTs using the \textit{PubMed} database:
\begin{verbatim}
("gene expression" OR "nucleotide*" 
OR "*omic*"  OR "genetic signature" 
OR "SNPs") AND (trial[Title/Abstract]) 
AND ((ffrft[Filter]) AND 
(randomizedcontrolledtrial[Filter]) 
AND (2019/5/1:2021/5/1[pdat]))
\end{verbatim}
This search, performed on June 2021, covers publication of RCTs between May 1st 2019 and May 1st 2021, with at least one of the terms: ``gene expression", ``nucleotide", ``omics", ``genetic signature" or ``SNPs", appearing in the title or abstract. We included full-text articles published in English. The search identified 174 papers which were screened for eligibility. 

After preliminary screening of titles and abstracts, eight reviewers (SC, TB, MJG, JN, LO, FW, SFW, JMSW) independently assessed the full text of relevant publications for final inclusion.

Papers were deemed eligible if they described RCTs that collected high-dimensional data. Here, data variables refer to biological variables collected at randomisation that could be used for comparing between the treatment arms or stratifying patients. For the purpose of this review, we adopted a flexible definition of high-dimensionality with respect to the number of variables. Specifically, we included studies containing at least 10 variables as they could benefit from methods suitable for high-dimensional data. An additional study that analysed 7 SNPs was included in this review as it used a multivariable approach.

We analysed the type of high-dimensional data (e.g.\ DNA, gene expression, etc.), the number of covariates used, purpose of collecting high-dimensional data, method of analysis of high-dimensional data, clinical area, and number of treatment arms. See the Supplementary Materials for the full summary of extracted data. 

\subsection{Results}
Out of the 174 papers returned, 100 (57.5\%) met the inclusion criteria. A summary of the  data extracted from included articles is given in Table \ref{tab:summary}. 

\begin{table}[h!] 
	\caption{Summary of extracted data. The denominator used to compute the percentages is 100 (number of eligible papers) unless specified. The most common answers appear in bold.  
		\label{tab:summary}}
	\centering
	\begin{adjustbox}{width=\columnwidth}
		\begin{threeparttable}
			\begin{tabular} {lll}\hline
				\Centering \bfseries Question & \Centering \bfseries Answer &  \Centering \bfseries $\bm{n}$ (\%) \\ \hline 
				\multirow{6}{100pt}{Type of high-dimensional data}
				& {\bf Gene expression} &  {\bf 70 (70\%)} \\ 
				& DNA  &  21 (21\%)\\
				& Metabolomic data & 1 (1\%) \\
				& Multiple data types\tnote{1} & 4 (4\%)\\
				& Proteomic data & 3 (3\%)\\
				& Questionnaire & 1 (1\%)\\ \hline
				\multirow{4}{100pt}{Number of covariates used} 
				& $<$10  & 1 (1\%)  \\
				& {\bf 10--100} & {\bf 41 (41\%)}\\
				& 101--1000 & 20 (20\%)\\
				& $>$1000 & 38 (38\%)\\
				\hline
				\multirow{9}{100pt}{Method of analysis\tnote{2}}
				& {\bf Univariable approach} &  {\bf 58 (36.3\%)}\\
				& Multivariable approach & 28 (17.5\%) \\
				& Gene sets and networks & 12 (7.5\%)\\
				& Cluster analysis & 18 (11.25\%)\\
				& Principal component analysis & 17 (10.6\%)\\
				& Penalised regression  & 5 (3.1\%)\\
				& Risk scores & 4 (2.5\%) \\
				& Not stated & 2 (1.25\%)\\
				& Other\tnote{3} & 16 (10\%)\\  \hline
				\multirow{5}{100pt}{Clinical area}
				& {\bf Oncology} & {\bf 30 (30\%)}\\
				& Chronic diseases & 28 (28\%) \\
				& Nutrition and ageing & 18 (18\%) \\
				& Cardiovascular diseases & 7 (7\%) \\
				& Other\tnote{4} & 17 (17\%)\\ \hline
				\multirow{3}{100pt}{Number of treatment arms}
				& {\bf 2} & {\bf 79 (79\%)}\\
				& 3 & 17 (17\%)\\
				& 4 & 4 (4\%) \\ \hline
			\end{tabular}
			\begin{tablenotes}
				\item[1] Questionnaires, omics data, biochemical characteristics and laboratory parameters.
				\item[2] The denominator used to compute these percentages is 160, because 42 (42\%) studies used multiple methods of analysis.
				\item[3] Functional analysis, Shannon entropy and Simpson index, significance analysis of microarrays, single sample predictor classifier, SVM.
				\item[4] HIV, malaria, mental health, neuropathy, ophthalmology.
			\end{tablenotes}
		\end{threeparttable}
	\end{adjustbox}
\end{table}

Most of the articles were for clinical trials in oncology (30\%) and various chronic diseases (28\%), including liver, kidney, rheumatic and respiratory diseases.  Other clinical areas included nutrition and ageing (18\%) and cardiovascular diseases (7\%).

The majority of articles (70\%) analysed gene expression data. The second most common type of data analysed was DNA data (21\%), including genome-wide SNP data. Five percent of the articles analysed metabolomic data, protein data and data from questionnaires. Four percent of the articles analysed multiple types of data. 

We divided the number of analysed covariates into four categories: ``$<$10", ``10--100", ``101--1000", and ``$>$1000". A large proportion of the analysed articles (41\%) had 10--100 covariates available for  analysis. A similar proportion of articles (38\%) used $>$1000 covariates in their analysis. Fewer studies (20\%) had 101--1000 covariates for the analysis, and one study had seven covariates, thus falling into the ``$<$10" category.

The methods used in the analyses and their advantages are summarised in Table \ref{tab:methodsummary}. 42\% of the studies used multiple methods of analysis. The most common analysis technique was a univariable analysis (36.3\%), followed by a multivariable analysis (17.5\%), cluster analysis (11.25\%) and PCA (10.6\%). Other methods that were reported  included: gene networks, multiple correspondence analysis, \citep{greenacre:blasius:2006} penalised approaches and risk scores,  Shannon entropy and Simpson index, \citep{shannon:1948,simpson:1949} significance analysis of microarrays, \citep{tusher:etal:2001} single sample predictor classifier,\citep{guinney:etal:2015} SVM.  

Most trials had two arms (79\%), followed by three-arm (17\%) and four-arm trials (4\%). In this review, we only analysed RCTs and therefore single-arm studies have been excluded. 

The purpose of collecting high-dimensional data varied substantially between trials and was often not reported clearly. For those trials where it was reported, categorisation of the reasoning proved challenging. Some trials used high-dimensional data as the (primary or secondary) outcome by analysing the effect of the intervention on gene expression, for example. In some cases, high-dimensional data was used to explore predictive biomarkers or to compare treatment arms. In other cases, the prognostic properties of the high-dimensional data were investigated, i.e.\ they did not compare the treatment arms but analysed the data as if it were observational.

\begin{figure} 
	\centering
	\includegraphics[scale = 0.4]{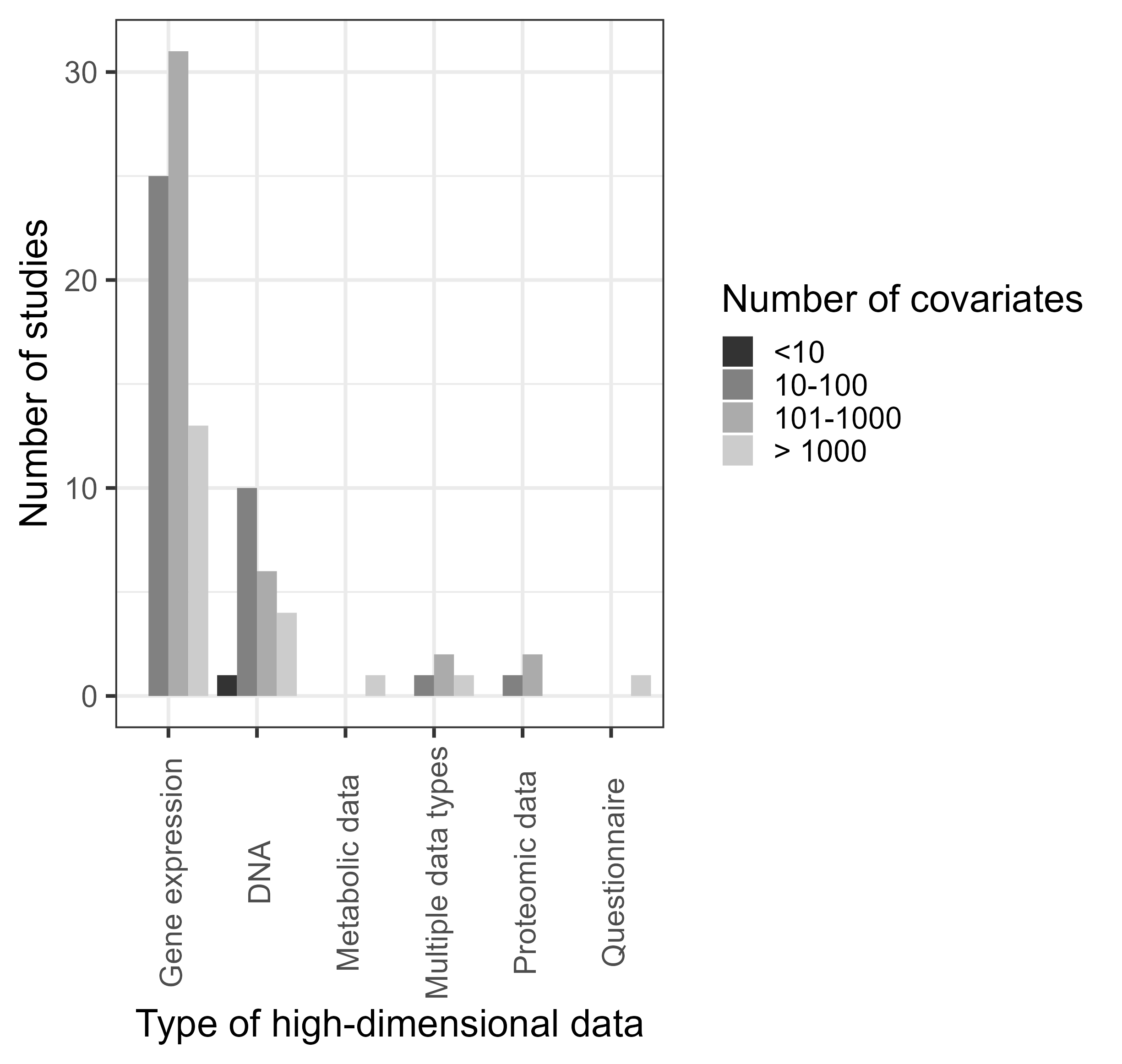} 
	\caption{Number of covariates per type of high-dimensional data.} \label{fig:ncov_type}
\end{figure}

\begin{figure} 
	\centering
	\includegraphics[scale = 0.4]{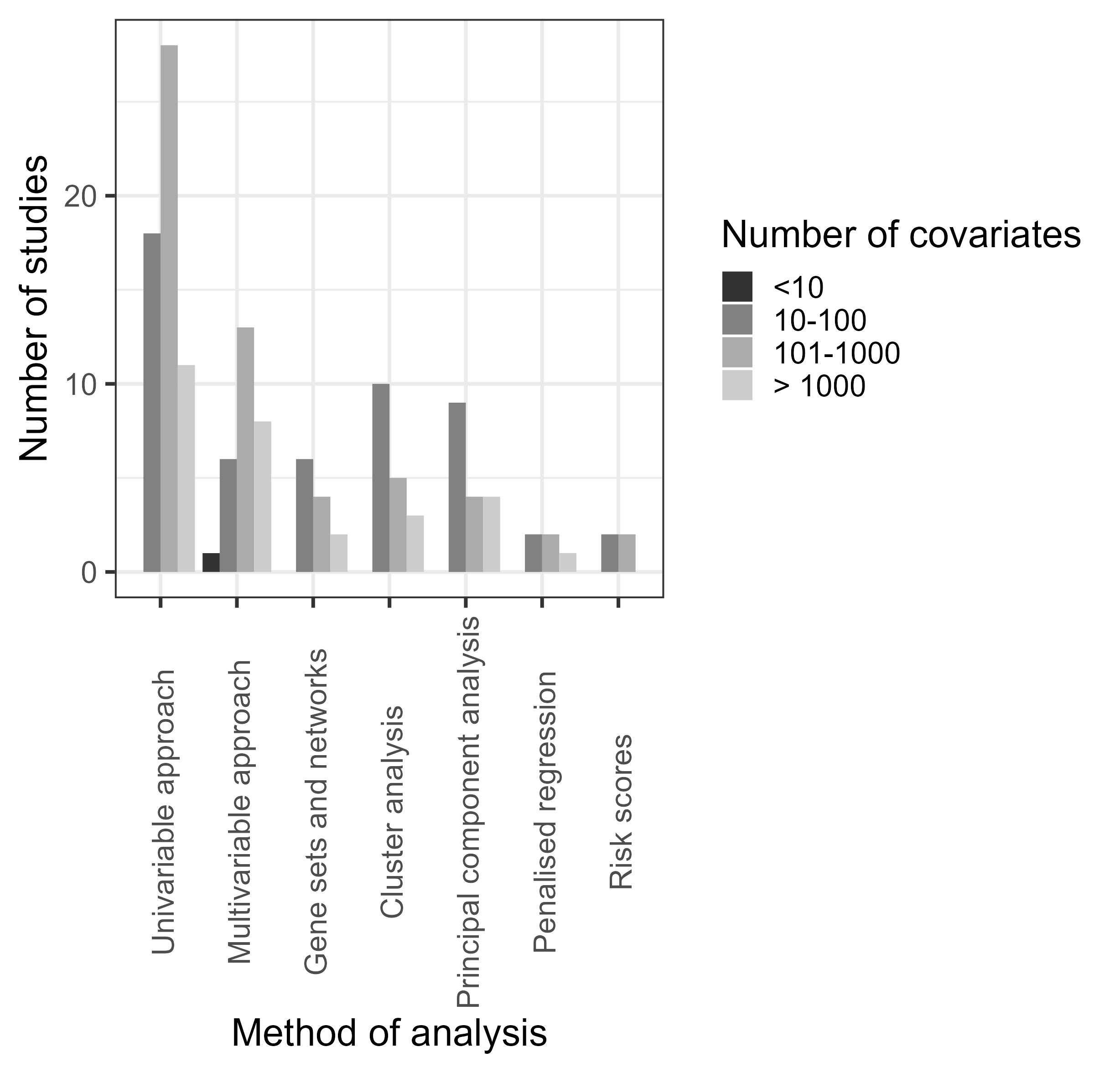} 
	\caption{Number of covariates per method of analysis.} \label{fig:ncov_method}
\end{figure}

\begin{figure}
	\centering
	\includegraphics[scale = 0.4]{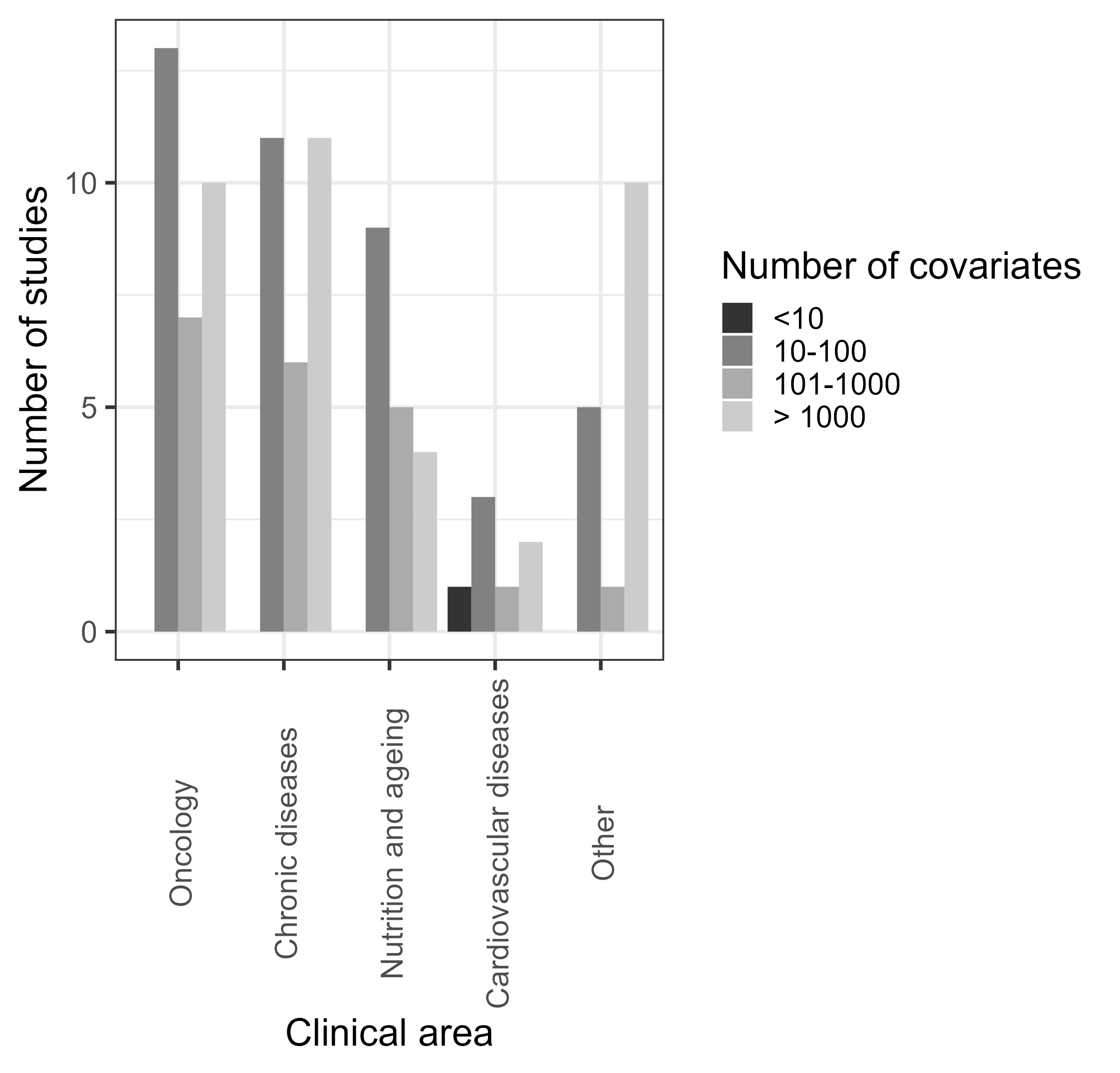} 
	\caption{Number of covariates per clinical area.} \label{fig:ncov_area}
\end{figure}

Figures \ref{fig:ncov_type}-\ref{fig:ncov_area} show the distribution of different types of data, methods of analysis and clinical areas, respectively, stratified by the number of covariates.
With regards to data types, most studies used gene expression or DNA and had 10-100 covariates, 100-1000 covariates or $>$1000 (Figure \ref{fig:ncov_type}). Regarding analysis methods, most studies using univariable and multivariable approaches utilised 101-1000 covariates, while gene sets and networks, clustering, and PCA most commonly used 10-100 covariates (Figure \ref{fig:ncov_method}).	In oncology, chronic diseases, and nutrition and ageing, the most common number of variables was 10-100; a substantial number of studies across all clinical areas analysed a larger number of covariates (101-1000 and $>$1000, Figure \ref{fig:ncov_area}).

\section{Discussion}
In this paper, we  provided an overview of methods for analysing high-dimensional data collected in clinical trials. We  also reviewed 100 recently published articles reporting RCTs that utilised high-dimensional data to identify which methods are typically used in practice. Although we focused on high-dimensional genetic data, the methods described could be applied to other types of high-dimensional data, such as questionnaires, imaging data or data from wearable technologies.

In our search, gene expression and DNA data were the most common data analysed, covering a combined total of 91\% of the high-dimensional data types included. A majority of the articles collected a large number of genetic data ($>$1000 variables), which  reflects the progress in  high-throughput technologies  and highlights the need for increased uptake of more sophisticated methods to utilise the high-dimensional data efficiently.

Although most of the trials we reviewed had two-arms, over 20\% had three or four arms. This reflects the additional complexity and challenges of utilising high-dimensional data in conjunction with multi-arm trials. Most clinical trials in this review were in the areas of oncology (30\%) and several chronic diseases (28\%). One of the challenges of trials for chronic diseases is learning how best to treat patients in the long-term. In particular, different treatments might be used for patients at different disease stages. Therefore, efficient methods that utilise changes in high-dimensional data over time are needed, for example methods that utilise longitudinal modelling.

Although we found some examples of more sophisticated methods being used to analyse high-dimensional data, the majority implemented straightforward approaches to examine interactions, such as univariable analysis. Methods such as machine learning, penalised approaches and risk scores appeared rarely in the analysis. For example, LASSO was seldom used despite being widely studied and having advantages. Adaptive signature design was not used. In some studies, high-dimensional data were measured, but only a small proportion of it was analysed. Therefore, there is strong potential for much more efficient use of high-dimensional data.

We investigated the distribution of the number of covariates across data types, methods of analysis and clinical areas. The number of covariates varied widely in each of these settings, highlighting  the need for developing methods that would be applicable to data of different orders of magnitude.   

High-dimensional data was collected for a variety of reasons, from being the primary outcome to identifying prognostic biomarkers  in exploratory analysis to investigating biological pathways. However, few studies used high-dimensional data to compare treatments or to identify predictive biomarkers, which highlights a gap and presents an opportunity to use the data more effectively. 

The limited use of sophisticated methods could be explained by perceived complexities and limitations of using high-dimensional data in clinical trials. Firstly, high-dimensionality of the data still requires \textit{a priori} knowledge of the disease mechanism, in the form of existing disease classification, to efficiently reduce the dimensionality of the data.\citep{theilhaber:etal:2020} Secondly, there may be a discrepancy between the signature constructed from genetic data and its biological meaning, which obscures the intuitive interpretation of high-dimensional data. For example, it has been found that a large number of breast cancer signatures constructed from a variety of gene sets do not explain the biological mechanism of the disease.\cite{manjang:etal:2021} In oncology, the most common field that collected high-dimensional data according to this review, this leads to genetic signatures being rarely used in clinical trials. It has been suggested that incorporating different types of omics data and using standardised methodology has the potential to make more effective use of high-dimensional data in clinical trials in order to improve patient outcomes. \citep{qian:etal:2021} In this review, we have only described the methods that were used in the analysed studies. Alternative methods, such as Bayesian classifiers, \citep{lampimen:vehtari:2001} also have the potential to analyse high-dimensional data in clinical trials.

In conclusion, although we only used a single database and limited timelines, we show that an increasing number of clinical trials are collecting high-dimensional data. Many of them  could benefit from implementing more sophisticated analysis methods, such as those outlined in this manuscript.  Further research is needed to make full use of the high-dimensional data collected in RCTs.

\appendix

\section{Appendix}
Consider a hypothetical randomised placebo-controlled clinical trial of $n$ participants with a normally distributed outcome $N(\mu_0, \sigma^2)$ for the control arm, and $N(\mu_1, \sigma^2)$ for the experimental arm, The number of participants in each group is the same ($n/2$).
We would like to test $H_0: \delta = 0$ where $\delta = \mu_1 - \mu_0$. A Wald statistic to test $H_0$ would be
\begin{equation*}
W = \frac{\hat \delta}{\sqrt{\frac{4\sigma^2}{n}}}.
\end{equation*}
For a two-sided $\alpha$ significance level, the sample size $n$ required for power $1-\beta$ is
\begin{equation*}
n = \frac{4 (Z_{1-\alpha/2} + Z_{1-\beta})^2\sigma^2}{\delta^2}.
\end{equation*}

Now suppose we have a binary biomarker that divides the population into biomarker-positive and biomarker-negative patients, with $r$ being the proportion of biomarker-positive patients. 
We assume that the treatment effect is $\delta_+ = \mu_{1+} - \mu_{0+}$ in biomarker-positive patients, and  $\delta_- = \mu_{1-} - \mu_{0-}$ in biomarker-negative patients. The treatment-biomarker interaction effect, $\delta_+ - \delta_-$, could be estimated by 
\begin{align*}
&\hat\delta_+ - \hat\delta_-  = (\hat\mu_{1+} - \hat\mu_{0+}) - (\hat\mu_{1-} - \hat\mu_{0-}) \\
& \sim N \left(\delta_+ - \delta_-, \frac{\sigma^2}{rn} + \frac{\sigma^2}{rn} +\frac{\sigma^2}{(1-r)n} +\frac{\sigma^2}{(1-r)n}\right).
\end{align*}
A Wald statistic to test $H_0: \delta_+ - \delta_- = 0 $ would be
\begin{equation*}
W_{int} = \frac{\hat\delta_+ - \hat\delta_- }{\sqrt{\frac{\sigma^2}{rn} + \frac{\sigma^2}{rn} +\frac{\sigma^2}{(1-r)n} +\frac{\sigma^2}{(1-r)n}}} = \frac{\hat\delta_+ - \hat\delta_-}{\sqrt{\frac{4\sigma^2}{nr(1-r)}}}.
\end{equation*}
For a two-sided $\alpha$ significance level, the sample size $n_{int}$ required for power $1-\beta$ is
\begin{equation*}
n_{int} = \frac{4 (Z_{1-\alpha/2} + Z_{1-\beta})^2\sigma^2}{(\delta_+ - \delta_-)^2r(1-r)}.
\end{equation*}
Thus, $n_{int} = \frac{n}{r(1-r)}$, i.e\ the sample size required to detect treatment-biomarker interaction increases by factor $\frac{1}{r(1-r)}$, with $\mathrm{min}  \left\{ \frac{r}{1(1-r)} \right\} = 4$ for $r = 0.5$. Therefore, the sample size for detecting the treatment-biomarker interaction is at least four times higher than the sample size needed to detect the main treatment effect.

\begin{dci}
The author(s) declared no potential conflicts of interest with respect to the research, authorship, and/or publication of this article.
\end{dci}

\begin{funding}
	
	This research was supported by the Medical Research Council (MR/S014357/1). JMSW, LO and SC are funded by the National Institute for Health and Care Research (NIHR301614).
\end{funding}

$\mathrm l\mathrm o\mathrm g\mathrm i\mathrm t\left(p_i\right)=\log\left[P\left(Y_i=1\right)/\left\{1-P\left(Y_i=1\right)\right\}\right]$

$\log\left\{h_i\left(t\right)/h_{0i}\left(t\right)\right\}$

$\begin{array}{c}{\widehat\delta}_+-{\widehat\delta}_-=\left({\widehat\mu}_{1+}-{\widehat\mu}_{0+}\right)-\left({\widehat\mu}_{1-}-{\widehat\mu}_{0-}\right)\\\sim N\left(\delta_+-\delta_-,\frac{\sigma^2}{rn}+\frac{\sigma^2}{rn}+\frac{\sigma^2}{\left(1-r\right)n}+\frac{\sigma^2}{\left(1-r\right)n}\right)\end{array}$

\bibliographystyle{SageV}
\bibliography{refs}

\end{document}